\documentclass[12pt]{article}

\usepackage{amssymb}
\usepackage{epsfig,color}
\usepackage{pstricks,graphicx,epsfig,color,amssymb,amsmath,amscd}
\usepackage{cite}
\usepackage{subcaption}
\usepackage[]{graphicx}
\usepackage{makeidx}

\newcommand{\be}{\begin{eqnarray}}
\newcommand{\ee}{\end{eqnarray}}
\newcommand{\rar}{\rightarrow}

\usepackage[]{caption}
\renewcommand\rho{\varrho}

\begin{document}

\begin{titlepage}
\title{Vacuum energy problem in {anti-de Sitter space}} 

\author{E.V. Arbuzova$^{a,b}$, A.D. Dolgov$^{b,c}$}

\maketitle
\begin{center}
$^a${Department of Higher Mathematics, Dubna State University, \\Universitetskaya ul. 19, Dubna 141983, Russia}\\
$^b${Department of Physics, Novosibirsk State University, \\Pirogova 2, Novosibirsk 630090, Russia}\\
$^c${Bogolyubov Laboratory of Theoretical Physics, Joint Institute for Nuclear Research,
Joliot-Curie st. 6, Dubna, Moscow region, 141980 Russia}


\end{center}

\vspace{1cm}
\begin{abstract}
We investigate the cosmological evolution of a universe with an initially large negative vacuum energy, corresponding to an anti-de Sitter (AdS) space-time. Such a universe inevitably contracts and collapses into a singularity. To prevent this collapse, we propose a dynamical mechanism of vacuum energy compensation based on a  scalar field non-minimally coupled to the curvature scalar. Numerical and analytical calculations demonstrate that the suggested mechanism completely compensates the gravitational impact of the vacuum energy, successfully driving the curvature to zero. As a result, the universe avoids the singularity and evolves into a power-law expansion, typical for a radiation-dominated cosmology.
\end{abstract}

\end{titlepage}

{
\section{Introduction \label{s-intro}}

The problem of the cosmological constant was born in 1917~\cite{lam-ein}, when Einstein introduced the so-called Lambda-term into the classical Hilbert-Einstein action of General Relativity to compensate for the gravitational attraction of usual matter and allow for a stationary universe. However, it soon became clear that such a mechanism is intrinsically unstable. Moreover, the discoveries by Friedmann~\cite{Fr22,Fr24} and Lema\^itre~\cite{lem} of the expansion of the universe persuaded Einstein to abandon the stationary hypothesis, later famously referring to the cosmological constant as his ``biggest blunder''~\cite{Gamow}. Today, however, the $\Lambda$-term is widely identified with the vacuum energy density, which acts as the dark energy driving the current accelerated expansion of the universe~\cite{wDE, SIB-DAD}.

Despite this phenomenological success, the underlying nature of the vacuum energy presents one of the most profound theoretical challenges. While astronomical observations indicate a tiny dark energy density $\rho_{DE} \sim 1 \text{ keV/cm}^3 \approx 10^{-47} \text{ GeV}^4$, quantum field theory predicts vacuum energy contributions that exceed this value by dozens of orders of magnitude~\cite{vac-Pauli, vac-YaBZ}. As discussed in detail in our recent paper~\cite{EA-AD-susy}, phenomena such as supersymmetry breaking or the formation of quark~\cite{cond-q} and gluon~\cite{cond-glu} condensates in the QCD vacuum yield enormous vacuum energy densities. 
For instance, the  gluon  condensate contribution alone is expected to be roughly $47$ orders of magnitude larger than the observed value. If uncompensated, these contributions would violently contradict the observed cosmological evolution. 
In light of this, we should admit that the idea of the $\Lambda$-term, alias the cosmological constant, was not Einstein's greatest blunder but a brilliant insight. Natural theoretical estimates demand it to be huge, so the problem is turned upside down: why it is so small but still probably non-zero.

Various theoretical models have been proposed to address this immense discrepancy, 
pioneered with the simple version of the paper~\cite{ad-nuff} and later developed e.g. in
~\cite{Kamenshchik:2018ttr, Barvinsky:2018lyi, Henke:2017cwv, Bengochea:2019daa, Panda:2025clg},
 see also references therein. Earlier attempts to dynamically reduce vacuum energy via a scalar field non-minimally 
 coupled to the curvature scalar $R$ were explored in~\cite{AD-MK-1, AD-MK-2, AD-MK-3, AD-FU}. However, in those models, 
 the transition to a canonical cosmology dominated by ordinary matter 
was not generally achieved, because it leads to the strong time variation  
of the gravitational coupling constant, $G_N \sim 1/t^2$~\cite{ad-nuff}, which is at odds with observational data.
The same result of time varying $G_N$ in the similar model was rediscovered recently in Ref.~\cite{melchiori}.

Moreover the resulting total
energy-momentum tensor was not proportional to the metric tensor ($T_{\mu\nu} \neq \Lambda g_{\mu\nu}$), 
and therefore the vacuum energy does not vanish, even asymptotically. 

In our previous work~\cite{EA-AD-susy}, we proposed a novel dynamical mechanism for vacuum energy compensation in the 
case of an initially positive vacuum energy, corresponding to a de Sitter (dS) space-time. By introducing a specific non-minimal 
coupling between the scalar field and the curvature scalar, we demonstrated that the initial exponential 
expansion is  driven to a canonical power-law evolution typical for the cosmology dominated by relativistic matter. 
It means that vacuum energy effectively becomes zero.

In the present paper, we apply this approach to the
universe with an initially negative vacuum energy, corresponding to an anti-de Sitter (AdS) space-time. A universe dominated by a 
negative cosmological constant inevitably faces a rapid deceleration of its expansion, followed by a contraction phase 
ultimately arriving to a catastrophic Big Crunch singularity. According to our suggestion, 
this collapse could be prevented by a
compensating scalar field which acts swiftly and with sufficient strength to neutralize the negative 
curvature before the universe reaches the point of no return. 

Here, we investigate the viability, dynamics, and natural limits of our compensation mechanism in the AdS regime. We explicitly 
analyze the competition between the characteristic timescales of the gravitational collapse and the scalar field evolution, and 
determine the conditions under which the universe can avoid the singularity and 
safely arrive to a stable Friedmann-like expansion.

\section{Vacuum Energy and the Curvature Scalar \label{s-vac}}

The cosmological constant $\Lambda$ is introduced into the classical Hilbert-Einstein action of General Relativity as follows:
\be
 S_\Lambda =  - \frac{M_{Pl}^2}{16 \pi} \int d^4 x \sqrt{-g} (R +\Lambda) + S_{M}, 
 \label{HE-action_original}
\ee
where $g_{\mu\nu}$ is the metric tensor, $g$ is its determinant, $R$ is the curvature scalar, $M_{Pl} = 1.22 \times 10^{19} \text{ GeV}$ is the Planck mass, and $S_M$ is the matter action. 

It is well known that the cosmological constant $\Lambda$ is physically equivalent to the vacuum energy density:
\be 
 T_{\mu\nu}^{(vac)} = \rho_{vac} g_{\mu\nu} = {M_{Pl}^2 \Lambda }/(16\pi).
 \label{t-mu-nu_vac}
\ee
Using this relation, the General Relativity action in an empty space-time in the presence of a non-zero $\Lambda$ can be rewritten as:
\be
 S_\Lambda = - \frac{M_{Pl}^2}{16 \pi} \int d^4 x \sqrt{-g} R - \rho_{vac}  \int d^4 x \sqrt{-g}.
 \label{HE-action}
\ee
The corresponding vacuum Einstein equations read:
\be
 \frac{M_{Pl}^2}{8 \pi} \left( R_{\mu\nu} -\frac{1}{2} g_{\mu\nu} R\right) \equiv \frac{M_{Pl}^2}{8 \pi} \, G_{\mu\nu} = T_{\mu\nu}^{(vac)}.
 \label{Ein-eq_vac}
\ee
Taking the trace of Eq.~(\ref{Ein-eq_vac}), we find the direct relation between the curvature scalar and the vacuum energy density:
\be
 R = -\frac{32\pi}{M_{Pl}^2}\,\rho_{vac}.
 \label{R-of-rhovac}
\ee
Note that $R > 0$ for an anti-de Sitter space-time, since $\rho_{vac}$ is  negative.

\section{Properties of pure anti-de Sitter expansion}

Let us first consider a purely anti-de Sitter universe, i.e., a universe possessing only negative vacuum energy without any matter. Such a universe is inherently homogeneous and isotropic, and its metric should be of the classical Friedmann-Lema\^itre-Robertson-Walker (FLRW) form:
\be
ds^2 = dt^2 - a^2(t) d{\vec r}^{\,2} .
\label{ds2}
\ee
Here we have assumed that the three-dimensional space is flat. This simplifying assumption is not of importance for the mechanism of vacuum energy compensation which we consider below.

In this metric, the curvature scalar $R$ is expressed through the Hubble parameter $H = \dot a/a$ via the relation:
\be
R = -6\left(\dot H + 2H^2 \right).
\label{H-thru-R}
\ee

In what follows, we will use dimensionless quantities normalized to $H_0$, where $H_0^2 = 8\pi |\rho_{vac}| / (3M_{Pl}^2)$:
\be
H =H_0 h,\,\,\, R =H_0^2 r, \,\,\, t= \tau /H_0.
\label{dim-less}
\ee
In this dimensionless form, Eq.~(\ref{H-thru-R}) turns into:
\be
 \frac{dh}{d\tau} + 2 h^2 = -\frac{r}{6} .
\label{dh-1}
\ee

The solution of Eq.~(\ref{dh-1}) for an empty space filled only with vacuum energy reads:
\be
h = -\sqrt{\frac{r_0}{12}} \tan\left[\sqrt{\frac{r_0}{3}}\, \left(\tau - \tau_0 \right)\right] .
\label{sol-r-const}
\ee
Recall that according to our assumption, the initial value of the curvature, $r_0$, is positive because the vacuum energy is supposed to be negative (see Eq.~(\ref{R-of-rhovac})). The evolution of the dimensionless Hubble parameter is illustrated in Figure~\ref{fig:ads_expansion}.

\begin{figure}[htbp]
    \centering
    \begin{minipage}{0.48\textwidth}
        \centering
        \includegraphics[width=\linewidth]{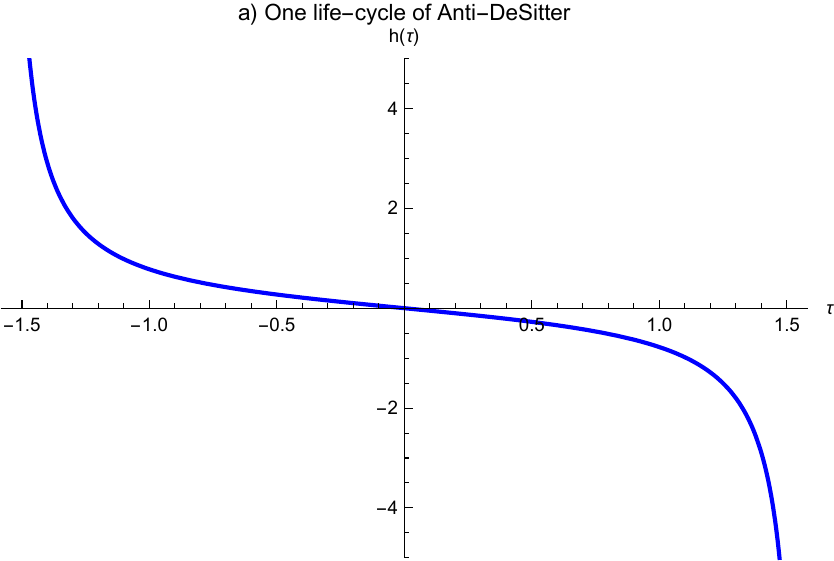}
    \end{minipage}\hfill
    \begin{minipage}{0.48\textwidth}
        \centering
        \includegraphics[width=\linewidth]{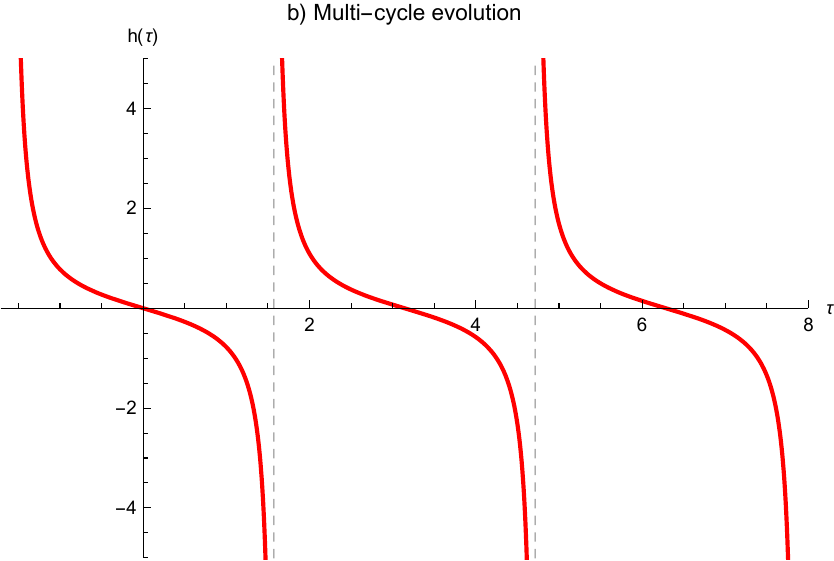}
    \end{minipage}
    
    \caption{Evolution of the dimensionless Hubble parameter \eqref{sol-r-const} as a function of time $\tau$: a) left panel: small values of $(\tau-\tau_0)$, which essentially covers the entire lifetime of the anti-de Sitter universe from its birth to expansion, and reflection back to a singularity; b) right panel: hypothetical evolution of several life-cycles of the anti-de Sitter universe, if it could somehow pass through the singularity.}
    \label{fig:ads_expansion}
\end{figure}

This result shows that a universe with a negative vacuum energy first expands (assuming that initially $h$ was positive), then 
stops when $h$ reaches zero, starts to contract when $h$ becomes negative, and finally collapses to a singularity. 
This is the typical behavior for an anti-de Sitter space-time.

 \section{GR equations with Lambda term and matter \label{s-lam-mttr}}

The Einstein equations, with the Lambda-term presented as vacuum energy, can be written as:
\be
 \frac{M_{Pl}^2}{8 \pi} \left( R_{\mu\nu} -\frac{1}{2} g_{\mu\nu} R\right) \equiv \frac{M_{Pl}^2}{8 \pi} \, G_{\mu\nu} = T_{\mu\nu}^{(vac)} +  T_{\mu\nu}^{(matt)}.
 \label{Ein-eq_matt}
\ee
Here, the energy-momentum tensor of vacuum has the form:
\be
 T_{\mu\nu}^{(vac)} = \rho_{vac} g_{\mu\nu}.
 \label{t-mu-nu-vac}
\ee
 In absence of matter $R= -32\pi \rho_{vac} /M_{Pl}^2$, see Eq. \eqref{R-of-rhovac}.

The matter component, $T_{\mu\nu}^{(matt)}$, is defined as the variation of the matter action over the metric tensor: 
\be
 T^{(matt)}_{\mu\nu} = \frac{2}{\sqrt{-g}}\, \frac{\delta S_M}{ \delta g^{\mu\nu}}.
 \label{TM-mu-nu}
\ee

To construct a dynamical mechanism for vacuum energy compensation, we introduce a real scalar field $\phi$. The action for this field is given by:
\be
 S_\phi = \int d^4 x \sqrt{-g} \left[ \frac{1}{2} g^{\mu\nu} \partial_\mu \phi\, \partial_\nu\phi - U(\phi,R) \right] ,
 \label{s-of-phi}
\ee
and the corresponding energy-momentum tensor has the standard form:
\be
 T_{\mu\nu}^\phi = (\partial_\mu \phi) (\partial_\nu \phi) - \frac{1}{2} g_{\mu\nu} \left[ g^{\alpha\beta} (\partial_\alpha\phi)( \partial_\beta \phi) - 2 U(\phi,R) \right] .
 \label{T-mu-nu-phi}
\ee

For a homogeneous field $\phi = \phi(t)$ in the FLRW metric, the variation of the action \eqref{s-of-phi} with respect to $\phi$ yields the following general equation of motion:
\be
 \ddot \phi + 3H \dot \phi +\frac{\partial U}{\partial \phi} = 0.
 \label{ddot-phi-R}
\ee

The first model attempting a dynamical reduction of vacuum energy for a  de Sitter space-time 
via the interaction of the curvature scalar 
$R$ with a scalar field was proposed in~\cite{ad-nuff}. 
Recall that for  de Sitter space-time vacuum energy is positive, and thus $R$ is negative, see Eq. (\ref{R-of-rhovac}).

It was assumed the coupling with the potential $U(\phi,R)$ chosen in the simplest form:
\be
 U = \frac{1}{2}\left( \beta R + m^2 \right)\phi^2.
 \label{U-of-phi-0}
\ee
It is easy to see that for positive $\beta$ the product  $\beta R < 0$ and hence if $|\beta R| > |m^2|$, 
the equation of motion \eqref{ddot-phi-R} in a de Sitter space-time possesses unstable solutions. If the Hubble
friction term $3 H\dot \phi$ can be neglected,  the solution rises exponentially with time, 
since at a constant curvature $R$ the square of the effective mass of the field $\phi$ becomes negative.
Anyhow it can be seen that the account of the Hubble friction does not eliminate rising of $\phi$ for sufficiently 
large $|{\beta}|$, a it has been shown in Ref.~\cite{EA-AD-susy}.

With rising $\phi$, its 
influence on the cosmological evolution can no longer be neglected. As can be easily verified, the initial exponential expansion, 
$a(t) \sim \exp(H_{vac} t)$, asymptotically transforms into a power-law one, $a(t) \sim t^\kappa$.

\section{Details of the compensation mechanism \label{s-comp-mech}}

To cure the problems of the simpler model described above and to effectively eliminate the vacuum energy, we introduce a more general coupling of the field $\phi$ with gravity. In the present work we closely follow our previous paper~\cite{EA-AD-susy} taking into account 
that the sign of the coefficient $\beta$ should be chosen opposite to compensate the  negative vacuum energy of the anti-de Sitter space.
We present the calculations here in some detail to make the paper self-contained.

We assume the following interaction Lagrangian density:
\be
 {\cal L}_f = \frac{1}{2}\left[ m^2 \phi^2 + \beta R Q (\phi) \right],
 \label{L-0}
\ee
where $Q (\phi) = \phi^2 f (\phi)$. The equation of motion for a homogeneous field $\phi(t)$ in the FLRW metric then takes the form:
\be 
 \ddot\phi + 3 H \dot \phi + m^2 \phi + \frac{1}{2} \beta R\, \partial_\phi Q = 0,
 \label{eq-2-phi}
\ee
where $\partial_\phi Q = \partial Q/\partial \phi$.

The energy-momentum tensor of $\phi$, defined via the variation of the action over the metric tensor \eqref{TM-mu-nu}, yields: 
\be \nonumber
 T_{\mu\nu}  &=& (\partial_\mu \phi) (\partial_\nu \phi) 
 -\frac{1}{2} g_{\mu\nu} \left[ g^{\alpha\beta} (\partial_\alpha\phi)( \partial_\beta \phi) - m^2 \phi^2 \right] \\
 && - \beta  Q(\phi)  \left( R_{\mu\nu} - \frac{1}{2} g_{\mu\nu} R \right) + \beta \left(D_\mu D_\nu - g_{\mu\nu} D^2 \right) Q(\phi),
 \label{t-mu-nu-c}
\ee
where $D_\mu$ is the covariant derivative. Using the equation of motion \eqref{eq-2-phi} and the identities $D_\mu Q = (\partial_\phi Q) \partial_\mu \phi$ and $D^2 Q = \partial^2_\phi Q \partial_\mu \phi \partial^\mu \phi +\partial_\phi Q D^2 \phi$, the trace of this energy-momentum tensor is found to be:
\be
 T^\nu_\nu =  -(\partial \phi)^2 ( 3\beta \partial_\phi^2 Q +1) + 2 m^2\phi^2 +\beta Q R + 3 \beta \left[(\partial_\phi Q)m^2\phi +\frac{1}{2}\beta R (\partial_\phi Q)^2 \right] .
 \label{trace-1}
\ee

As a consistency check, let us consider the special case where $Q = \phi^2$. The trace \eqref{trace-1} reduces to:
\be
 T^\nu_{\nu} = -(6\beta+1)( \partial_\mu \phi)( \partial^\mu \phi) + \beta (6\beta+1) R \phi^2 + 2 (1+3 \beta) m^2 \phi ^2. 
 \label{trace-tc}
\ee
This is a well-known result. Note that for a massless field ($m = 0$) and the specific conformal coupling $\beta = -1/6$, the trace of the energy-momentum tensor identically vanishes.

To execute the dynamical compensation mechanism in our scenario, we introduce a more complicated specific function $\bar Q(\phi,M_0,k)$:
\be 
 \bar Q(\phi,M_0,k) = \phi^2 (1 + \sigma \phi^2/M_0^2)^k ,
 \label{bar-Q}
\ee
with $M_0$, $k$, and $\sigma = \pm 1$ being constant parameters. In the present work, to compensate for the negative vacuum energy of the anti-de Sitter space-time, we choose $\sigma = 1$, $\beta = -1$, and assume the scalar field is massless ($m = 0$).

The derivatives of $\bar Q$ over $\phi$ are given by:
\be
 \partial_\phi {\bar Q } &=&\frac{2k \sigma  \phi^3 (1 + \sigma \phi^2/M_0^2)^{k-1}}{M_0^2} + 2 \phi (1 + \sigma \phi^2/M_0^2)^k,\label{Q'}\\
 \partial^2_\phi {\bar Q} &=& \frac{4 ( k-1) k \sigma^2 \phi^4 (1 + \sigma \phi^2/M_0^2)^{k-2}}{M_0^4} + \nonumber \\
 &&\frac{ 10 k \sigma \phi^2 (1 + \sigma \phi^2/M_0^2)^{ k-1}}{M_0^2} + 2 (1 + \sigma \phi^2/M_0^2)^k . 
 \label{Q'-Q''}
\ee

Finally, taking the trace of the Einstein equations \eqref{Ein-eq_matt} and combining it with \eqref{trace-1}, we find the 
algebraic relation expressing the curvature scalar $R$ through $\phi$ and its first derivative:
\be
 &&R\left(\beta \bar Q + \frac{3 \beta^2 (\partial_\phi \bar Q )^2}{2} +\frac{M_{Pl}^2}{ 8\pi}  \right) =\nonumber\\
 &&( \partial \phi)^2 ( 3\beta \partial_\phi^2 {\bar Q} +1) - 2 m^2\phi^2  
 -3 \beta  m^2\phi (\partial_\phi {\bar Q} )  - 4\rho_{vac} - \tilde{T}_\nu^\nu,
 \label{R-of-phi}
\ee
where $(\partial \phi)^2 = g^{\mu\nu} (\partial_\mu \phi) (\partial_\nu \phi)$, and $\tilde T^\nu_\nu$ is the trace of the energy-momentum tensor of other possible forms of matter (which vanishes for relativistic matter).

\section{Numerical calculations \label{s-numerical}}

There are the following two differential equations that govern the evolution of $\phi$ and the Hubble parameter:
\be
&&\ddot\phi + 3 H \dot \phi + m^2 \phi + \frac{1}{2} \beta R\, \partial_\phi Q  = 0,\\
&&\dot H + 2 H^2 = - R/6,
\label{sys-eqs}
\ee
where $R$ is a known function of $\phi$ and its first derivative given by Eq.~(\ref{R-of-phi}).
If $\beta = 0$, field $\phi $ is massless and  matter is relativistic, the curvature scalar evidently is a positive constant, equal to: 
\be
R(\beta = 0) = - \frac{32 \pi}{M^2_{Pl}}\, \rho_{vac} .
\label{R-beta-0}
\ee

To proceed further, it is convenient to introduce dimensionless variables:
\be 
&&\tau = t H_0,\, \varphi = \phi/H_0,\, h = H/H_0,\, \frac{d}{dt} = H_0 \frac{d}{d\tau},\nonumber\\
&&R = r H_0^2,\, \rho_{vac} = - \lambda H_0^4,\, \bar Q = H_0^2 q,\,  M_0  = H_0 \mu .
\label{dim-less2}
\ee
Here $H_0$ is a normalization constant. For definiteness, we fix it by the condition 
$H_0^2 = 8\pi|\rho_{vac}|/(3M_{Pl}^2)$, where $\rho_{vac}< 0$ is the original large negative vacuum energy of the anti-de Sitter space. We introduce a  positive dimensionless parameter $\lambda > 0$ to represent the magnitude of the vacuum energy,  $\lambda = |\rho_{vac}| / H_0^4$. 
In what follows, we denote the derivative over $\tau$ by a prime: $df/d\tau \equiv f'$.

The dimensionless function $q = \bar Q/H_0^2$ is expressed through the dimensionless field $\varphi$ as:
\be
q = \varphi^2 \left( 1 + \sigma \varphi^2/\mu^2\right)^k.
\label{q1}
\ee
The derivatives of $\bar Q$ \eqref{Q'} and \eqref{Q'-Q''} turn into:
\be
&&\frac{\partial_\phi {\bar Q}}{ H_0}  \equiv q_1
 = \frac{2 k\sigma \varphi^3 (1 + \sigma {\varphi^2}/{\mu^2})^{k-1}}{\mu^2} + 2 \varphi \left(1 + \sigma {\varphi^2}/{\mu^2}\right)^k,\\ \label{q1_deriv}\nonumber\\
&&\partial^2_\phi {\bar Q} \equiv q_2 = \frac{4 ( k-1) k \sigma^2\varphi^4 (1 + \sigma{\varphi^2}/{\mu^2})^{k-2}}{\mu^4} + \nonumber\\
&&\frac{ 10 k \sigma \varphi^2 (1 + \sigma{\varphi^2}/{\mu^2})^{ k-1}}{\mu^2} + 2 \left(1 + \sigma {\varphi^2}/{\mu^2}\right)^k .
\ee

Finally, we come to the following two differential equations governing the cosmological evolution:
\be 
&& h' + 2 h^2 = -\frac{r}{6} \label{dh},\\
&& \varphi'' + 3 h \varphi' + \beta r q_1/2 = 0,
\label{d2phi}
\ee
where the dimensionless curvature $r$ is determined by Eq.~\eqref{R-of-phi}. 

In what follows, we assume that the mass of the field $\phi$ is zero ($m = 0$) and that the trace of the energy-momentum tensor of other kinds of matter also vanishes, since it is supposed to be relativistic ($\tilde T_\nu^\nu = 0$). Using our normalization condition, the Planck mass term can be substituted as $M_{Pl}^2/(8 \pi H_0^2) = \lambda/3$. Accounting for the negative vacuum energy, $\rho_{vac} \sim (-\lambda)$, the curvature $r$ takes the following dimensionless form:
\be
r = \frac{( 3\beta q_2 +1) (\varphi')^2 + 4 \lambda }
{ 3\beta^2 q_1^2/2 + \beta q + \lambda/3 } .
\label{r0}
\ee

Equations \eqref{dh} and \eqref{d2phi}, with $r$ given by Eq.~\eqref{r0}, are solved numerically with the parameters chosen to compensate for the negative vacuum energy: $\sigma=1$ and $\beta=-1$. In Fig.~\ref{fig:ads_evolution}, the evolution of the dimensionless functions $h(\tau)$, $\varphi(\tau)$, $\varphi'(\tau)$, and $r$ is presented for the  magnitude of the dimensionless vacuum energy $\lambda = 10^2$. The figure details four distinct stages of the evolution during different time intervals: small $\tau$ (panel a), intermediate $\tau$ (panel b), large $\tau$ (panel c), and very large $\tau$ (panel d).

\begin{figure}[htbp]
    \centering
    \begin{subfigure}[b]{0.48\textwidth}
        \centering
        \includegraphics[width=\textwidth]{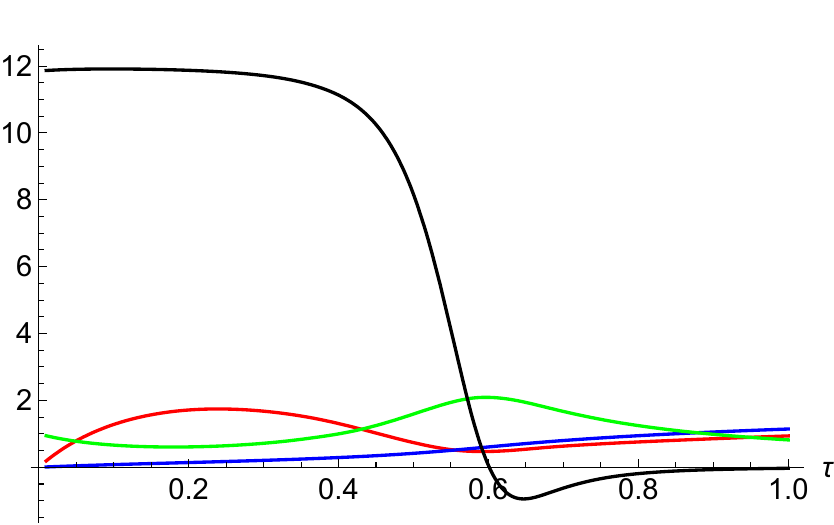}
        \caption{Small $\tau$}
        \label{fig:small_tau}
    \end{subfigure}
    \hfill
    \begin{subfigure}[b]{0.48\textwidth}
        \centering
        \includegraphics[width=\textwidth]{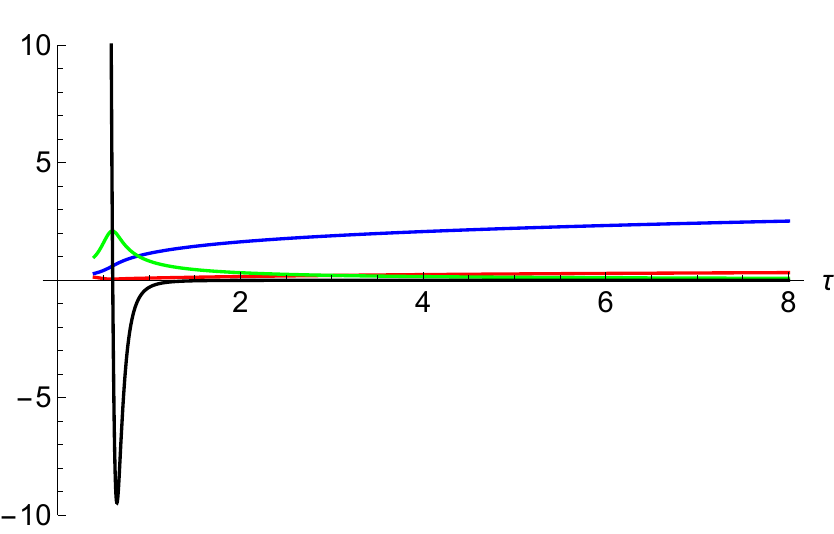}
        \caption{Intermediate $\tau$}
        \label{fig:int_tau}
    \end{subfigure}
        \vspace{0.5cm} 
    \begin{subfigure}[b]{0.48\textwidth}
        \centering
        \includegraphics[width=\textwidth]{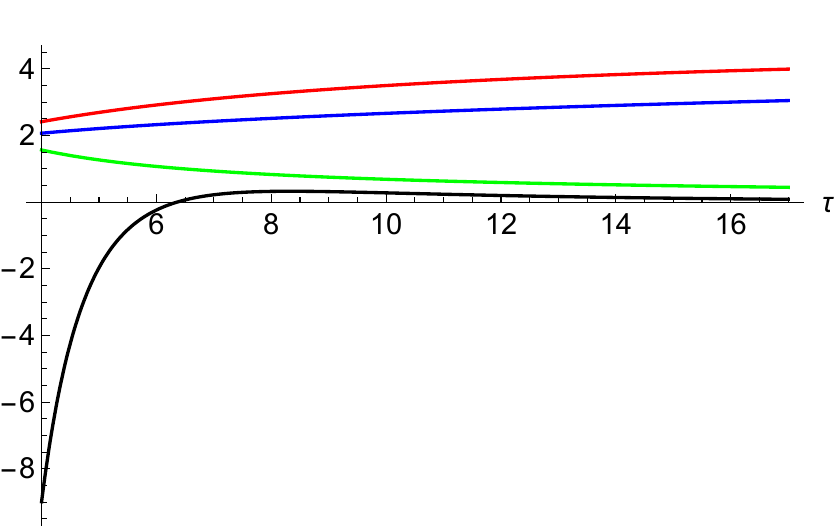}
        \caption{Large $\tau$}
        \label{fig:large_tau}
    \end{subfigure}
    \hfill
    \begin{subfigure}[b]{0.48\textwidth}
        \centering
        \includegraphics[width=\textwidth]{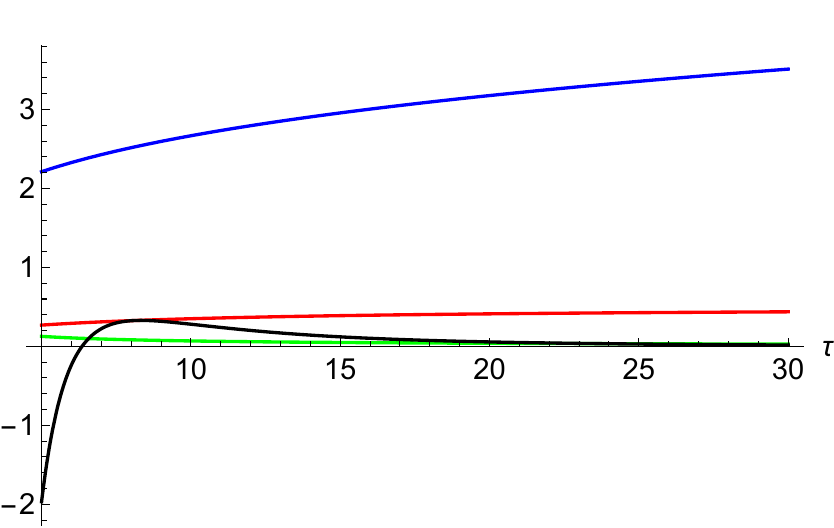}
        \caption{Very large $\tau$}
        \label{fig:very_large_tau}
    \end{subfigure}  
    \caption{Numerical solutions for the cosmological evolution in anti-de Sitter space with initial parameters $\lambda = 100$, $\beta = -1$, $k = 3$, and $\mu_f = 1$. The panels display the dimensionless Hubble parameter multiplied by time $\tau h$ (red lines), the scalar field $\varphi$ (blue lines), its derivative $\varphi'$ (green lines), and the dimensionless curvature $r$ (black lines) across different time intervals. To ensure visual clarity, the functions are scaled by specific factors in each panel: 
    (a) Small $\tau \in [0.01, 1]$: $10\tau h$, $\varphi$, $\varphi'$, $r$; 
    (b) Intermediate $\tau \in [0.4, 8]$: $\tau h$, $\varphi$, $\varphi'$, $10r$; 
    (c) Large $\tau \in [4, 17]$: $10\tau h$, $\varphi$, $10\varphi'$, $10^6 r$; 
    (d) Very large $\tau \in [5, 30]$: $\tau h$, $\varphi$, $\varphi'$, $10^6 r$.}
    \label{fig:ads_evolution}
\end{figure}

If $\lambda$ is not too large (in this case, $\lambda = 100$), our compensation mechanism successfully stabilizes the dynamics. As a result, we arrive at a positive and decreasing Hubble parameter, scaling roughly as $\sim 1/\tau$, alongside a decreasing curvature that asymptotically goes down to zero. Note that the curvature briefly becomes positive, which, according to Eq.~\eqref{dh-1}, further drives the decrease of $H$.

For a much larger vacuum energy, e.g., $\lambda = 10^3$, the field $\varphi$ is not strong enough, and the curvature remains negative. The Hubble parameter also remains negative, which corresponds to the contraction of the universe, as is typical for the usual anti-de Sitter space-time. In this case, $H$ tends to minus infinity, signifying the singularity of the contracting universe in the process of normal anti-de Sitter evolution, as shown in Fig.~\ref{fig:num_10_3}.}

\begin{figure}[htbp]
    \centering
    \begin{subfigure}[b]{0.48\textwidth}
        \centering
        \includegraphics[width=\textwidth]{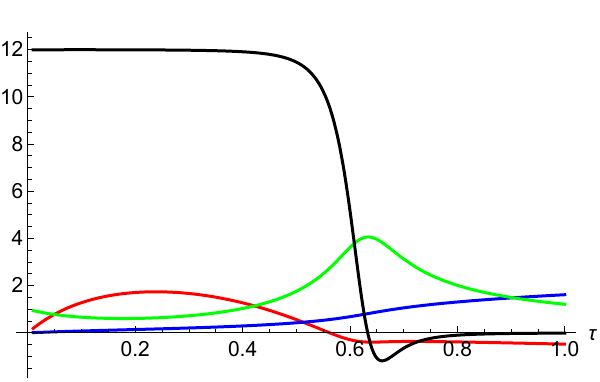}
        \caption{Small $\tau$}
        \label{fig:lam1000_small}
    \end{subfigure}
    \hfill
    \begin{subfigure}[b]{0.48\textwidth}
        \centering
        \includegraphics[width=\textwidth]{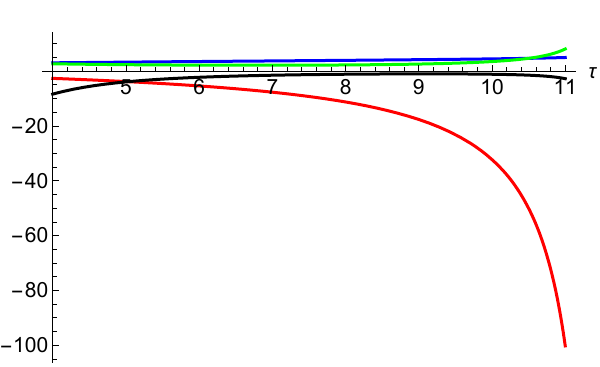}
        \caption{Large $\tau$}
        \label{fig:lam1000_large}
    \end{subfigure}
    
    \caption{ Cosmological evolution for an increased initial vacuum energy $\lambda = 10^3$ (with $\beta = -1$, $k = 3$, and $\mu_f = 1$). 
    The panels display the dimensionless functions: $10\tau h$ (red lines), $\varphi$ (blue lines), $\varphi'$ (green lines), and 
    $r$ (black lines) across different time intervals. To ensure visual clarity, some functions are scaled in each panel: 
    (a) Small $\tau$: $10\tau h$, $\varphi$, $\varphi'$, $r$; 
    (b) Large $\tau$: $10\tau h$, $\varphi$, $10\varphi'$, $10^6 r$. 
    Unlike the $\lambda = 100$ case, the compensation mechanism fails here, and the universe collapses into a singularity.}
    \label{fig:num_10_3}
\end{figure}

Consequently, the universe collapses into a Big Crunch singularity faster than the compensation mechanism can complete its work and stabilize the curvature. Therefore, our model predicts the existence of a critical vacuum energy threshold, $|\lambda_{crit}|$. If the initial negative vacuum energy exceeds this critical value, the dynamic compensation with $\beta = -1$ fails to prevent the ultimate collapse of the universe.

\begin{figure}[htbp]
    \centering
    \begin{subfigure}[b]{0.48\textwidth}
        \centering
        \includegraphics[width=\textwidth]{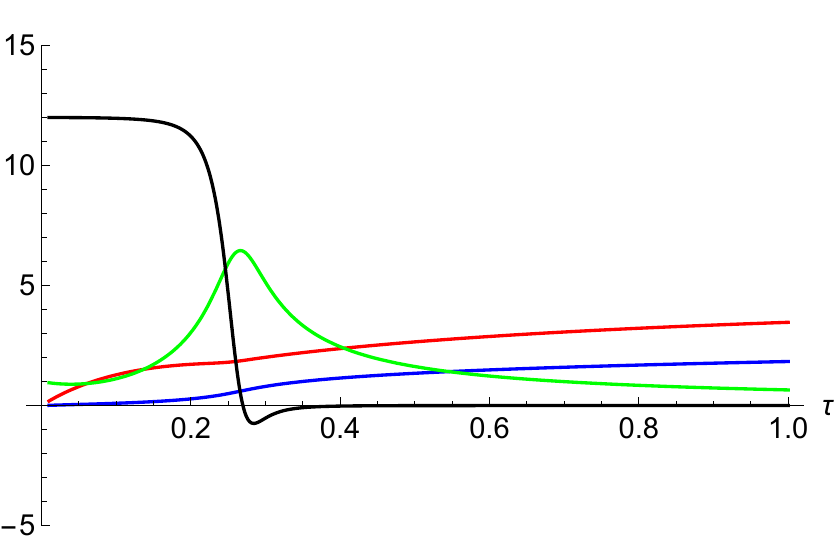}
        \caption{Small $\tau$}
        \label{fig:lam10000_small}
    \end{subfigure}
    \hfill
    \begin{subfigure}[b]{0.48\textwidth}
        \centering
        \includegraphics[width=\textwidth]{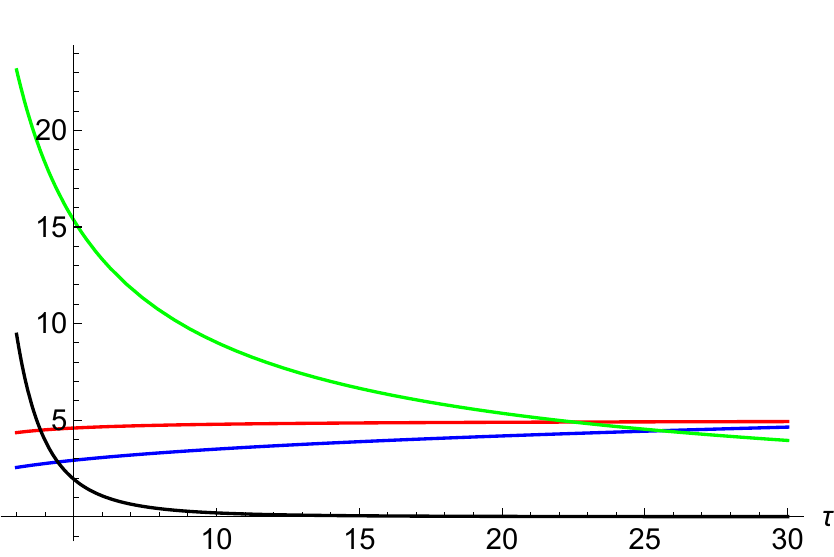}
        \caption{Large $\tau$}
        \label{fig:lam10000_large}
    \end{subfigure}
    
    \caption{Cosmological evolution in anti-de Sitter space for a huge initial dimensionless vacuum energy $\lambda = 10^4$ and an enhanced scalar field coupling $\beta = -10$ (with $k = 3$ and $\mu_f = 1$). The panels display the dimensionless Hubble parameter multiplied by time $\tau h$ (red lines), the scalar field $\phi$ (blue lines), its derivative $\phi'$ (green lines), and the dimensionless curvature $r$ (black lines). To ensure visual clarity, some functions are scaled by specific factors in each panel: 
    (a) Small $\tau \in [0.01, 1]$: $10\tau h$, $\phi$, $\phi'$, $r$; 
    (b) Large $\tau \in [3, 30]$: $10\tau h$, $\phi$, $10^2\phi'$, $10^7 r$.}
    \label{fig:extreme_vacuum_compensation}
\end{figure}

To overcome this failure, we increase the magnitude of the coupling parameter $\beta$ to enhance the compensating mechanism. Specifically, by setting it an order of magnitude larger ($\beta = -10$), we have found that the mechanism successfully drives the curvature to zero and prevents the singularity. Correspondingly, the anti-de Sitter evolution turns into a power-law expansion. In other words, vacuum energy is completely compensated, and we arrive at a canonical cosmology dominated by relativistic matter. The results of the numerical calculations    for $\lambda = 10^4$ and $\beta = -10$  are presented in Fig.~\ref{fig:extreme_vacuum_compensation}.

Equations \eqref{dh} and \eqref{d2phi} with $r = 0$ are trivially solved analytically, leading to the asymptotic results:
\be
h(\tau) \rar [2(\tau+\tau_0)]^{-1},\,\,\,  \varphi' \rar C  \tau^{-3/2},\,\,\, \varphi \rar const, 
\label{sol-analyt}
\ee
which agree remarkably well with the numerical calculations. Evidently, $H \sim 1/2t$ corresponds to the canonical radiation-dominated cosmology.

\section{Conclusion \label{s-concl}}

The model suggested in this paper efficiently compensates the effect of any original vacuum energy down to zero and leads to a resulting realistic cosmology governed by relativistic matter. 
Indeed, as demonstrated in Figs.~2 and 4, the value of the curvature scalar asymptotically vanishes. According to the standard relation for an arbitrary homogeneous and isotropic metric:
\be
R = -6 (\dot H + 2H^2), 
\label{R-of-H}
\ee
the asymptotic vanishing of $R$ leads to the Hubble parameter evolving as
\be
H = \frac{1}{2(t+t_0)},
\label{h-of-t}
\ee
and correspondingly, to the scale factor rising as $a(t) \sim t^{1/2}$, which is typical for a cosmological model dominated by relativistic matter.

It is important to emphasize that the physical vacuum energy itself does not disappear anywhere. Instead, its enormous gravitational impact is dynamically neutralized by the interaction of the curvature scalar with the scalar field. As a result of this compensation, the universe safely enters a standard power-law expansion regime, evolving exactly as if the vacuum energy were completely absent.

As the next steps, we need to include non-relativistic matter into the model and check if the transition to a matter-dominated cosmology could be successfully achieved. Another related problem is the possibility of describing cosmological dark energy within the proposed framework. Presumably, it can be realized by introducing a more complicated dependence on the curvature scalar $R$, analogously to the known description of dark energy by modified gravity through an $F(R)$ generalization of General Relativity. This is left for future studies.


\section*{Acknowledgments}
This work  was supported by state funding for neutrino physics FSUS-2025-0019.

\end{document}